\documentclass[twocolumn,secnumarabic,amssymb, nobibnotes, prb]{revtex4-2}
\usepackage{graphicx}
\usepackage{subfigure}
\usepackage{bm}
\usepackage{amsmath}
\usepackage{lineno}
\usepackage{color}
\begin{document}
\title{Rectification of plasma-powered active matter into macroscopic work}

\author{Ting-yu Yao$^{1}$}
\author{Shuo Wang$^{1}$}
\author{Shao-peng Li$^{1}$}
\author{Ming-hang Zhao$^{1}$}
\author{Bao-quan Ai$^{2}$}
\email[Email:]{aibq@scnu.edu.cn}	
\author{Ya-feng He$^{1,3}$}
\email[Email:]{heyf@hbu.edu.cn}

\affiliation{$^1$College of Physics Science and Technology, Hebei University, Baoding 071002, China\\
$^2$Key Laboratory of Atomic and Subatomic Structure and Quantum Control (Ministry of Education), Guangdong Basic Research Center of Excellence for Structure and Fundamental Interactions of Matter, School of Physics, South China Normal University, Guangzhou 510006, China\\
$^3$Hebei Research Center of the Basic Discipline for Computational Physics, Baoding 071002, China.}

\date{\today}
\begin{abstract}
   Plasmas sustain strong energy and momentum flux, yet this activity is typically treated as a dissipative loss channel rather than a usable resource, and its conversion into macroscopic work remains challenging. Here we demonstrate a plasma-powered active engine by coupling self-propelled micromotors to ratchet rectification. Dielectric microspheres in the plasma sheath spontaneously develop asymmetric surface charging and undergo a Quincke rotational instability, forming fast micromotors that extract energy directly from the plasma. Their stochastic impacts are rectified by a sawtooth rotor into a directed angular-momentum flux that drives steady rotation, while an outer asymmetric gear organizes the active bath into coherent circulation to amplify torque transfer. Operating in an inertia-relevant regime, the engine achieves orders-of-magnitude enhancements in both power output and end-to-end efficiency compared with liquid-phase active engines, and remains functional down to the single-micromotor limit. More broadly, our results establish a general route to rectify nonequilibrium plasma activity for macroscopic work.
\end{abstract}


\maketitle
\indent Converting dissipated energy into useful, predictable work is a central challenge across physics, biology, chemistry, and engineering \cite{Liu,Kim,Fan,Donelan,Yang}. In nonequilibrium systems, such conversion can occur without violating the second law when appropriate symmetry breaking elements are introduced \cite{Hanggi,Bricard,LiuS,Feynman,Roeling,Roche}. Two representative elements are active agents and Feynman ratchet. Active agents--ranging from motile bacteria to synthetic asymmetric micromotors-continuously draw energy from their surroundings to sustain persistent motion \cite{Bechinger,Scholz,Keta}. However, their motion is often effectively unbiased, so the harvested energy is not readily converted into useful work at the scale of interest. Ratchet mechanisms, by contrast, rectify unbiased fluctuations into directional transport by exploiting broken spatial symmetry, and have been demonstrated in diverse nonequilibrium settings \cite{Skaug,Pumm,He,Yao}. Combining active agents as the harvesting element with a ratchet architecture as the rectifying element naturally yields an ``active engine'' that transduces environmental dissipation into useful mechanical work \cite{Leonardo,Vizsny,Maggi,Ai,Jerez}. Yet most existing active engines operate in liquids, where strong viscous damping enforces overdamped dynamics [low Reynolds (Re) and P\'eclet (Pe) numbers]. This limits particle speeds (typically $\sim$1--100 $\mu$m\,s$^{-1}$) and suppresses impulsive momentum transfer to macroscopic loads, thereby constraining both single-particle power and overall conversion efficiency. 

\indent Gaseous plasma holds tremendous energy and offers a distinct regime for active matter: its weakly damped, inertia-relevant environment allows particles to reach high speeds and deliver strong impulsive momentum transfer. Here, we experimentally demonstrate an active engine in plasma that harvests plasma energy and converts it into mechanical output. Dielectric microspheres undergoing a Quincke rotational instability move with speeds up to \(\sim 10\ \mathrm{cm\,s^{-1}}\), behaving as micromotors. A collection of these micromotors forms an active bath that exhibits a dynamic competition between self-propulsion and diffusion. With rectification by an outer gear, the active bath drives a rotor to rotate persistently at high speed up to 4.2~rad\,s\(^{-1}\), achieving efficient utilization of stochastic energy. As a result, both the mechanical power output and the end-to-end energy-conversion efficiency exceed those of liquid-phase active engines by several orders of magnitude.
\begin{figure*}[htbp]
	\begin{center}
		\includegraphics[width=15.1cm,height=8cm]{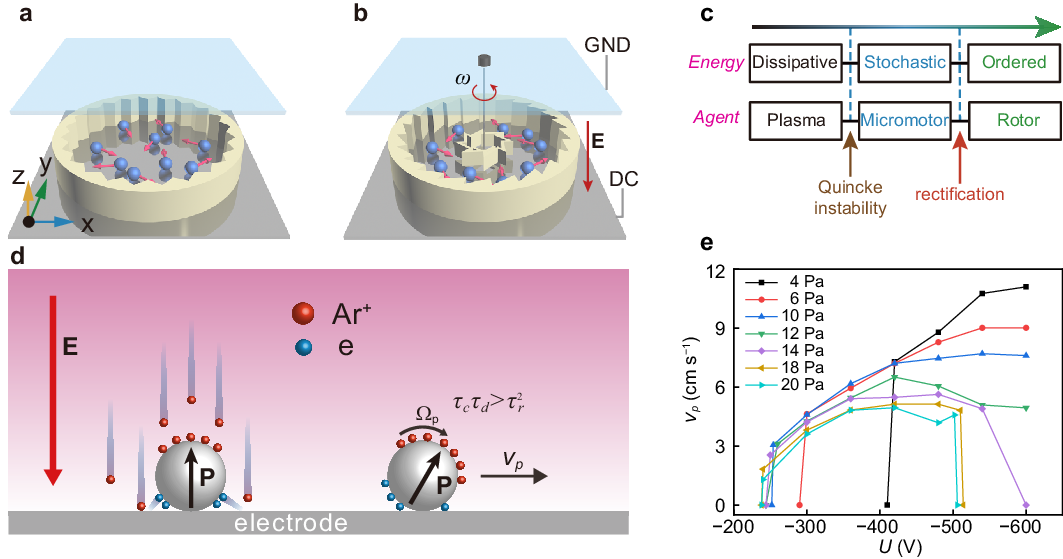}
		\caption{\textbf{Plasma-powered active engine via micromotor rectification.}
			{\bf a}, Stochastic state without rectification. In a symmetric outer gear, plasma-activated microspheres become self-propelled micromotors that form an active bath but exhibit stochastic motion.
			{\bf b}, Rectified state with work output. An asymmetric outer gear breaks spatial symmetry, converting random micromotor motion into coherent circulation that generates a net angular-momentum flux and drives persistent rotation of a magnetically suspended ratchet rotor.
			{\bf c}, Energy conversion pathway. Plasma dissipation is converted into micromotor kinetic energy and rectified into macroscopic mechanical work.
			{\bf d}, Micromotor origin. Sheath-driven ion streaming induces asymmetric surface charging of microspheres, generating an effective dipole \(\mathbf{P}\) that leads to a Quincke-type rotational instability; rolling converts rotation into fast translation.
			{\bf e}, High-speed micromotor dynamics. Self-propelled speed as a function of discharge voltage at different gas pressures, reaching up to \(\sim 10~\mathrm{cm\,s^{-1}}\), characteristic of a weakly damped, inertia-relevant regime. Particle radius $r_d=500~\mu$m.}
	\end{center}
	\vspace{-0.6cm}
\end{figure*}

\noindent \textbf{Results}\\ 
\textbf{System overview.} We realize a plasma-powered active engine by coupling a bath of self-propelled micromotors to a ratchet rotor (Fig.~1a,b). Monodisperse resin microspheres (radius \(r_d\sim\) hundreds of \(\mu\)m) are placed on the lower electrode in a DC argon glow discharge. When the plasma is ignited (Methods), the microspheres become asymmetrically charged and roll spontaneously and randomly on the smooth electrode surface, forming a tunable active bath of micromotors (Fig.~2a, Supplementary Video 1). The micromotor activity is controlled by the discharge conditions (typically gas pressure \(p=4\)--20~Pa and applied voltage \(U=-240\) to \(-660\)~V; Fig.~1e), consistent with continuous energy uptake from the nonequilibrium plasma.

\indent To convert this stochastic motion into macroscopic work, we introduce a magnetically suspended eight-tooth rotor into the active bath (Fig.~1b; Methods). Frequent collisions between the self-propelled micromotors and the rotor teeth transfer angular momentum, driving persistent rotation of the rotor (Fig.~3, Supplementary Video 2). To further enhance the angular-momentum flux, we place an asymmetric outer gear around the rotor. This outer gear rectifies the random micromotor trajectories into a coherent circulating flow, amplifying the net torque delivered to the rotor and substantially improving energy-conversion performance (Supplementary Video 3). Overall, plasma dissipation is transduced into directed rotor motion via a two-stage process: energy uptake by micromotors followed by geometry-controlled conversion into macroscopic work (Fig.~1c).

\begin{figure}[htp]
	\begin{center}
		\includegraphics[width=8cm,height=7.6cm]{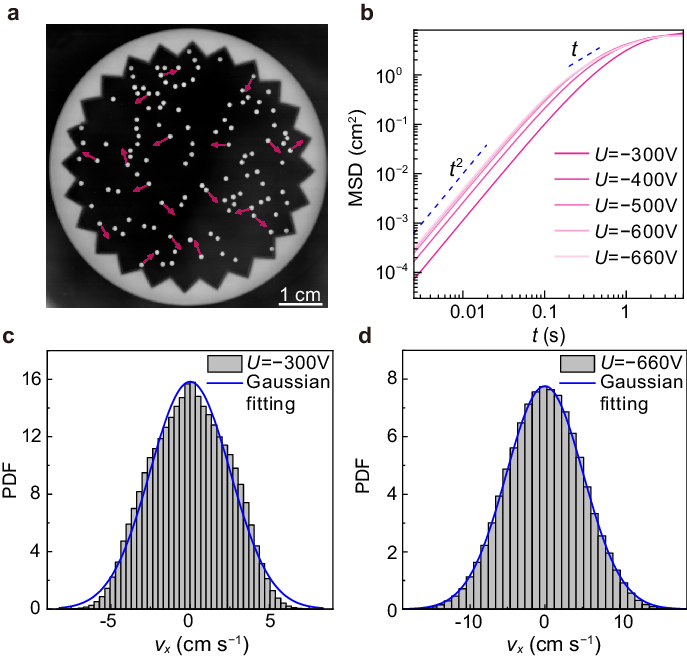}
		\caption{
			{\bf a}, Spatial distribution of micromotors confined within a symmetric outer gear, forming an isotropic active bath.
			{\bf b}, Mean-squared displacement of micromotors for different applied voltages. The dynamics exhibits the transition from short-time ballistic to long-time normal diffusive scaling. 
			{\bf c,d}, A statistical crossover from non‑Gaussian velocity distributions at low voltage (\textbf{c}) to Gaussian forms at high voltage (\textbf{d}). Particle radius $r_d=500$ \(\mu\)m. Gas pressure $p= 6$ Pa.}
	\end{center}
	\vspace{-0.6cm}
\end{figure}
\begin{figure*}[htp]
	\begin{center}
		\includegraphics[width=16cm,height=7.5cm]{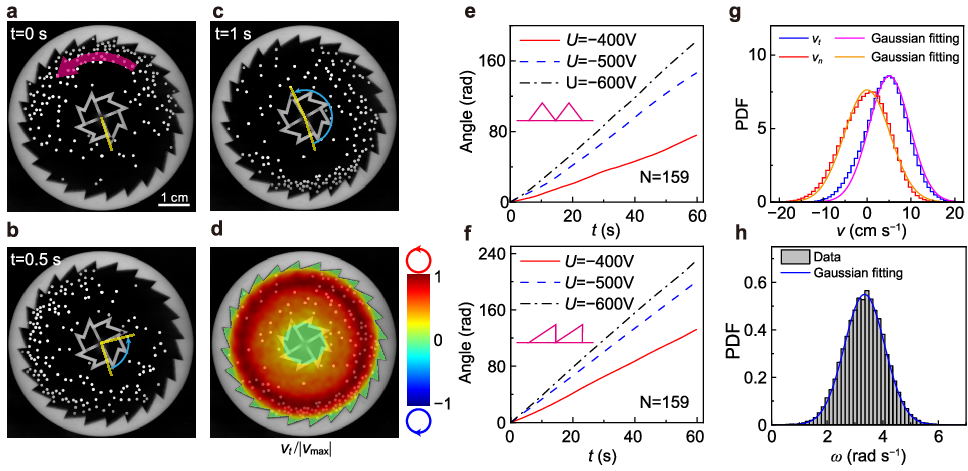}
		\caption{\textbf{Geometry-controlled rectification converts stochastic micromotor motion into mechanical work.} 
			{\bf a–c}, Time sequence showing persistent counterclockwise rotation of the central eight-tooth sawtooth rotor driven by micromotor impacts. An asymmetric outer gear rectifies initially random micromotor motion into coherent circulation, enhancing directional angular-momentum transfer (outer and inner sawtooth orientations are optimized for maximal rectification). 
			{\bf d}, Particle-image velocimetry (PIV) map of the rectified micromotor bath, revealing a robust counterclockwise circulating flow. 
			{\bf e,f}, Rotor angular displacement $\theta(t)$ driven by stochastic micromotor motion without rectification (\textbf{e}) and by rectified circulating micromotors (\textbf{f}). Rectification suppresses fluctuations and markedly increases the mean rotation rate. 
			{\bf g}, Probability distributions of radial velocity $v_n$ and tangential velocity $v_t$ components of micromotors in the rectified state, showing a finite mean tangential velocity and near-zero radial drift. 
			{\bf h}, Distribution of rotor angular velocity, demonstrating stable high-speed rotation under rectified driving. Discharge conditions: \(p=6\)~Pa, \(U=-500\)~V. Particle radius $r_d= 500$~\(\mu\)m.}
	\end{center}
\end{figure*}
\noindent \textbf{Self-propelled mechanism.} The dielectric microsphere harvests the dissipative energy from surrounding plasma when it becomes self-propelled within the plasma sheath. In the non-electroneutral sheath region above the lower electrode (Fig.~1d), the downward sheath electric field \(\mathbf{E}\) accelerates positive ions toward the electrode, producing asymmetric surface charging of the microsphere through two effects: (i) ion streaming past the microsphere favours positive-charge accumulation on the upper hemisphere; and (ii) ion impact on the lower electrode triggers secondary electron emission, enhancing electron collection on the lower hemisphere. Owing to the dielectric mismatch between the microsphere and the surrounding plasma, the characteristic charge‑relaxation time of the microsphere is much longer than that of the ambient plasma. Together, these effects generate a persistent charge imbalance across the microsphere surface, resulting in a metastable dipole moment \(\mathbf{P}\) antiparallel to the sheath electric field.

\indent Once the Quincke instability of this induced dipole is triggered, the microsphere begins to spin steadily \cite{Quincke}. A small perturbation that destabilizes the antiparallel dipole configuration gives rise to an electric torque \(\mathbf{P}\times\mathbf{E}\) that can overcome the neutral drag torque \(-\Omega_P/\tau_d\), driving sustained rotation. The spinning microsphere then rolls on the electrode surface, converting rotation into translation and functioning as a self-propelled micromotor with its translational motion perpendicular to the local electric field. The spin rate \(\Omega_P\) is governed by three characteristic timescales (see Supplementary Information Sec. I),
\begin{eqnarray}
	\Omega_P&=&\frac{1}{\tau_c}\sqrt{\frac{\tau_c \tau_d}{\tau_r^2}-1},
\end{eqnarray}
where \(\tau_c\) is the charging time, \(\tau_d\) the drag (retardation) time, and \(\tau_r^2=-J/(|\mathbf{P}\times\mathbf{E}|)\) the rotational time constant (with \(J\) the rotational inertia). Here, the onset condition \(\tau_c\tau_d/\tau_r^2>1\) defines the threshold for the Quincke instability, which corresponds to a critical experimental voltage for the onset of microsphere spin (Fig.~1e). Under the present discharge conditions, the measured translational speed first increases and then decreases as the applied voltage is raised.

\noindent \textbf{Micromotor dynamics.} An individual plasma-powered micromotor exhibits long, nearly ballistic ``flights'' on the smooth electrode surface (Supplementary Video 4) owing to weak damping (\(\gamma\sim 2.13\times10^{-7}\) kg\,s$^{-1}$) and a high P\'eclet number (Pe \(\sim10^4\)–\(10^6\))—in marked contrast to the overdamped run-and-tumble dynamics typical of liquid-phase systems \cite{Karani,Mano}. The measured maximum instantaneous speed reaches up to \(\sim10\)~cm\,s\(^{-1}\) in the present experiments, at least two orders of magnitude higher than typical liquid-phase active colloids \cite{ZhangB,ZhangZ,Garza}. From the short-time ballistic regime of the mean-squared displacement (MSD), well described by \(\mathrm{MSD} = 4D_t t + v_p^2 t^2\), we extract self-propelled speeds \(v_p = 5.1~\mathrm{cm\,s^{-1}}\) (at $-$300~V, 6~Pa). Due to the weak damping and low thermal noise, the self-propelled speed closely matches the mean instantaneous speed over a wide range of discharge conditions.

\indent A collection of micromotors naturally forms an active matter far from equilibrium by confining many micromotors within a symmetric outer gear (Figs.~1a,2a, and Supplementary Video 2). In this confined geometry, randomly moving micromotors undergo frequent particle--particle and particle--wall collisions, including both soft Yukawa repulsion and hard-contact events depending on relative speed. Using a central-collision method and the measured relative velocities of colliding pairs, we infer an effective particle charge \(Q\approx7\times10^7 ~e\) and a screening length \(\lambda\approx0.7\) mm (Supplementary Information, Sec.~\uppercase\expandafter{\romannumeral2}). The statistical properties of the micromotor ensemble reveal a dynamic competition between directed self-propulsion and stochastic diffusion (Supplementary Information, Sec.~\uppercase\expandafter{\romannumeral3}). Experimentally, the distribution of the instantaneous velocity component along the $x$ direction \(P(v_x)\) exhibits a pronounced voltage dependence (Fig.~2c,d). At lower voltages (e.g., $-300~\mathrm{V}$, Fig.~2c), the diffusion coefficient remains relatively low ($D_{\text{eff}}=3.18$ cm$^2$/s), and the self-propulsion speed amplifies the probability density near the characteristic speed (at \(v_{px}=3.3\)~cm\,s\(^{-1}\)), yielding a non-Gaussian distribution with symmetric enhancements on the shoulders of the distribution profile. At high voltage (e.g., \(-660\)~V; Fig.~2d), diffusion is strongly enhanced ($D_{\text{eff}}=13.22$ cm$^2$/s) and dominates over self-propulsion, so the distribution approaches a Gaussian form. Despite these changes in instantaneous statistics, long-time transport shows a robust crossover from short-time ballistic motion (\(\mathrm{MSD}\propto t^{2}\)) to long-time normal diffusion (\(\mathrm{MSD}\propto t\); Fig.~2b).

\noindent \textbf{Rectification and mechanical output.} To convert the energy of the active bath of randomly moving micromotors into mechanical output, we introduce a central sawtooth rotor (a Feynman ratchet) magnetically suspended within this bath. The sawtooth geometry of the rotor breaks azimuthal symmetry of the system. Frequent collisions between randomly moving micromotors and the asymmetric sawteeth drive slow but persistent anticlockwise rotation of the rotor and thus mechanical output. In this random-collision-driven regime, the rotation rate is low (2.44~rad\,s\(^{-1}\) at $-$500~V; Fig.~3e). Despite the relatively low mean rotation rate, individual collisions transfer substantial momentum owing to the inertia-relevant regime (Re \(\approx1\)--4). As a result, micromotors can drive a rotor orders of magnitude more massive than a single micromotor (\(m_R=0.44\)~g versus \(m_p\approx4\times10^{-4}\)~g), with single impacts producing angular steps up to \(\sim0.3^\circ\). Importantly, the system remains operational down to the single-micromotor limit: a single micromotor can drive the rotor through approximately one revolution within 4 minutes (Supplementary Video 5).

\begin{figure}[htp]
	\begin{center}
		\includegraphics[width=6.7cm,height=8.7cm]{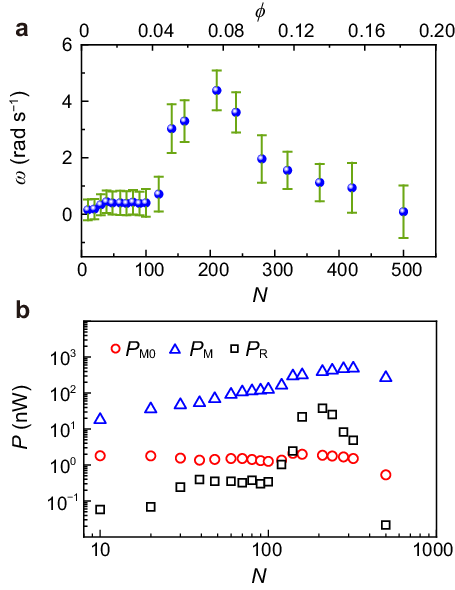}
		\caption{\textbf{Optimal loading and power.} 
			{\bf a}, Rotor rotation rate as a function of micromotor number \(N\) (areal fraction \(\phi\)), exhibiting a maximum at intermediate \(\phi\) due to the competition between enhanced collision-driven angular-momentum transfer and crowding-induced motility suppression. Error bars, 1 s.d.
			{\bf b}, Mechanical power of the rotor (squares), the micromotor ensemble (triangles), and a single micromotor (circle) as a function of \(N\). The finite single-particle contribution highlights that the engine remains operational down to the single-micromotor limit, while collective effects maximize total power at intermediate loading. Discharge conditions: \(p=6\)~Pa, \(U=-500\)~V. Particle radius $r_d= 500$ \(\mu\)m.}
	\end{center}
	\vspace{-0.6 cm}
\end{figure}
\indent To amplify the net angular-momentum flux to the rotor, we further rectify micromotor motion using an outer gear with 24 asymmetric sawteeth (Fig.~3a--c; Supplementary Video 3). Repeated scattering from these sawteeth generates a collective anticlockwise circulation of micromotors (Fig.~3d), characterized by a finite mean tangential velocity \(v_t\approx6.5\)~cm\,s\(^{-1}\) and a near-zero mean radial component (Fig.~3g) under \(p=6\)~Pa and \(U=-500\)~V. The rectified circulation also increases the mean particle speed by \(\sim10\%\) compared with the unrectified case, consistent with a collective enhancement. This rectification-induced circulation greatly increases the net angular-momentum flux delivered to the central rotor, producing rapid and steady rotation with strongly suppressed fluctuations in \(\theta(t)\) (Fig.~3f). Under these conditions, the rotor reaches 3.33~rad\,s\(^{-1}\) at $-$500~V (Fig.~3h), demonstrating that converting disordered micromotor motion into an ordered circulating flow substantially boosts mechanical output.

\textbf{Optimal loading and efficiency.} The rotor speed increases with micromotor activity, controlled by the applied voltage (Fig.~3e,f), and depends non-monotonically on the micromotor number $N$ (equivalently, the areal fraction \(\phi\); Fig.~4a). At low \(\phi\), increasing \(N\) enhances the collision rate between micromotors and the rotor sawteeth, thereby increasing angular-momentum transfer and accelerating rotation. At high \(\phi\), however, crowding suppresses micromotor motility and reduces the net torque on the rotor, eventually halting rotation. Under our conditions, an intermediate areal fraction (\(N\approx210\), \(\phi\approx0.07\)) maximizes the rotation rate, reaching 4.2~rad\,s\(^{-1}\) at \(p=6\)~Pa and \(U=-500\)~V.
 
 \indent We estimate the mechanical power and energy-conversion efficiencies of the system (Fig.~4b). The rotor power is calculated as \(P_\text{R}=\Gamma \omega^2\), where \(\omega\) is the measured rotor angular speed and \(\Gamma\sim2\times10^{-9}~\mathrm{m^2\,s^{-1}}\) is an effective drag coefficient, yielding \(P_\text{R}\sim 38\)~nW at optimal loading (Fig.~4b, squares). The mechanical power associated with a single micromotor is estimated as \(P_\text{M0}=\gamma v_p^2\), which is on the order of 1~nW for typical values \(r_d=500~\mu\mathrm{m}\) and \(v_p\sim10\)~cm\,s\(^{-1}\) (Fig.~4b, circles), which is approximately six orders of magnitude larger than that of bacteria \cite{Leonardo}. The total mechanical power of the micromotor ensemble, \(P_\text{M}=\sum P_\text{M0}\), increases with \(N\) and reaches up to \(500\) nW (Fig.~4b, triangles). The micromotor-to-rotor conversion efficiency, defined as \(\eta_{\text{M}\rightarrow\text{ R}}=P_\text{R}/P_\text{M}\), reaches \(\sim10\%\) under optimal loading, highlighting efficient conversion of bath-level mechanical activity into rotor work in the weakly damped plasma environment. The electrical input power for generating plasma \(P_{\mathrm{input}}=UI\) spans \(\sim10^{-2}\) to a few watts over our operating window ($-$300 to $-$660~V and 0.1 to 8~mA), implying an end-to-end efficiency \(\eta=P_\text{R}/P_{\mathrm{input}}\) of order \(10^{-8}\) (typically \(10^{-9}\)--\(10^{-7}\)), which is at least two orders of magnitude larger than that of light-driven rotor \cite{Vizsny}.                

\indent Energy conversion in this active engine proceeds in two stages: plasma \(\rightarrow\) micromotor kinetic energy \(\rightarrow\) rotor work. The end-to-end efficiency is therefore set jointly by plasma-to-micromotor uptake and micromotor-to-rotor coupling. The former can be increased by tuning the discharge voltage and loading more micromotors, which boosts the total activity and mechanical power available in the bath. The latter can be improved by tailoring the ratchet geometry to enhance micromotor speeds and to increase both the rate and the directional bias of micromotor--sawtooth impacts, thereby amplifying the net angular-momentum flux delivered to the rotor. Low pressure is particularly favourable because reduced neutral drag increases micromotor speed and rotor rotation, and can simultaneously lower the discharge power in our operating window, providing a practical route to higher overall efficiency. 

\indent In summary, we demonstrate that a nonequilibrium plasma can be harnessed as a practical energy source for an active engine: sheath-activated micromotors form a stochastic angular-momentum bath, and geometry-controlled rectification converts this activity into steady macroscopic work. By organizing random impacts into a directed angular-momentum flux---and further enhancing it via collective circulation---the design couples microscopic plasma-driven motion to a robust mechanical output and remains effective down to the single-micromotor driving limit. Operating in a weakly damped, inertia-relevant regime, the engine delivers power output and end-to-end energy-conversion efficiency that exceed those of liquid-phase implementations by orders of magnitude. Beyond energy converting, the same rectification principle could enable programmable transport and sorting of charged microparticles in low-temperature plasmas. This work establishes a general design principle for converting microscopic nonequilibrium fluctuations into macroscopic work in weakly damped charged media.\\

\noindent \textbf{Methods}\\
\noindent \textbf{Gear fabrication.}\\ 
The outer gears and the central rotor are fabricated by 3D laser printing (lateral resolution \(\sim10~\mu\)m) using a photopolymer resin (Somos$^{\circledR}$ Imagine 8000). The printed structures are cleaned to remove uncured resin and fully hardened prior to use. 

The symmetric and asymmetric outer gears have identical dimensions except for their tooth geometry. Both gears are 10.5 mm high, with inner and outer radii of 25 and 30 mm, respectively, and contain 24 sawteeth of depth 3 mm. The hollow central rotor has a wall thickness of 1 mm and a height of 5.5 mm. It contains eight asymmetric sawteeth with long and short side lengths of 6 and 3 mm, respectively. The maximum and minimum rotor radii are 9.5 and 6.6 mm, respectively. A cross-shaped support (height 1 mm) connects four nonadjacent sawteeth and contains a central hole for mounting the suspension needle.

\noindent \textbf{Rotor suspension.}\\
A thin pivot needle is vertically mounted through the center of the rotor and magnetically coupled to a miniature permanent magnet attached to the lower surface of the upper electrode, thereby minimizing mechanical friction. The rotor--electrode gap is maintained below \(200~\mu\mathrm{m}\), smaller than the microsphere diameter, preventing particle trapping while ensuring efficient momentum transfer during micromotor--rotor collisions. The lower electrode is mirror-polished to provide a low-friction surface for micromotor motion.\\

\noindent \textbf{Experimental procedure.}\\
 Experiments are performed in a DC argon glow-discharge chamber consisting of two parallel electrodes separated by 4.5 cm. Macroporous adsorbent resin microspheres (Cool Chemistry; Amberlite XAD4, relative size dispersion \(<10\%\)) are deposited on the lower electrode prior to plasma ignition. A weak argon flow (2--10 SCCM) is maintained throughout the experiments to stabilize the discharge while minimizing flow-induced perturbations and neutral-drag effects.

After the gas pressure and discharge voltage are set, the plasma is allowed to reach a steady state. Particle motion is recorded in real time through the transparent upper ITO electrode using a top-view camera (Mikrotron Eosens 4CXP, 1696$\times$1710 pixels, 10$-$400 fps). For each operating condition, videos are acquired over sufficiently long duration to characterize both the short-time ballistic and long-time diffusive regimes of particle motion.\\

\noindent \textbf{Parameter estimation.}\\
The mechanical power of a single micromotor is estimated as \(P_\text{M0}=\gamma v_p^2\), where \(\gamma=6\pi\eta r_d\) is the effective viscous drag coefficient, \(\eta=2.26\times10^{-5}~\mathrm{Pa\,s}\) is the viscosity of argon under the experimental conditions. This expression corresponds to the power dissipated against Stokes drag and provides a conservative estimate of the mechanical power in the weakly damped regime.
The rotor power is calculated as \(P_\text{R}=\Gamma \omega^2\), where \(\omega\) is the measured angular velocity and \(\Gamma\) is the effective rotational drag coefficient. The coefficient \(\Gamma\) is independently determined from rotor-relaxation measurements by recording the angular deceleration after switching off micromotor driving (or in control experiments without particles) and fitting the decay to a linear drag law, \(J\dot{\omega}=-\Gamma\omega\) over the experimentally accessible velocity range.\\

\noindent \textbf{Data availability}\\
All data that support the findings of this study are provided in this paper and the Supplementary Information. Source data are provided with this paper.

\begin{flushleft}
	\textbf{Acknowledgements}
\end{flushleft}
This work was supported by the National Natural Science Foundation of China (Grants No. 12475203, No. 12275064 and No. 12475036), the National Key R\&D Program of China (Grant No. 2025YFF0512000), and the China Manned Space Program (Grant No. CMSS-2025-5-P-004).

\begin{flushleft}
	\textbf{Author contributions}\\
\end{flushleft}
Y.F.H. and B.Q.A. designed the experiment and model. T.Y.Y., S. W. and S.P.L. performed the experiments, T.Y.Y. and M.H.Z. conduced the formulation of the model. All authors analyzed the data and wrote the manuscript.

\begin{flushleft}
	\textbf{Competing Interests}\\
\end{flushleft}
The authors declare that they have no competing financial interests.
\begin{flushleft}
	\textbf{Correspondence}\\
\end{flushleft}
Correspondence and requests for materials should be addressed to Ya-feng He or Bao-quan Ai.
\begin{flushleft}
	\textbf{Additional Information}\\
\end{flushleft}
Supplementary Information is available for this paper.

\end{document}